\documentstyle[12pt,epsf]{l-aa}
\newcommand{\bm}[1]{\mbox{\it \bf #1}}

\begin{document}
\thesaurus{12.03.4;12.04.1;12.12. }        
\title{Gravitational  Lagrangian Dynamics  of  Cold Matter  Using
  the  Deformation Tensor.}   
\author{Edouard Audit  \&  Jean-Michel Alimi  }  
\institute{$^1$   Laboratoire d'Astrophysique   Extragalactique  et de
  Cosmologie, CNRS URA 173, Observatoire de Paris-Meudon, Meudon 92195
  CEDEX, France} 
\date{} 
\maketitle
\markboth{Cosmological  gravitational dynamics  using  the deformation
  tensor}{}

\def\etal{{\it et al. \/}}
\def\eg{{e.g.,\ }}
\def\etc{{etc.\ }}
\def\ie{{i.e.,\ }}

\begin{abstract}
  In this paper we present a new local Lagrangian approximation to the
  gravitational dynamics of cold matter.  We  describe the dynamics of
  a  Lagrangian fluid  element   through   only  one  quantity,    the
  deformation tensor.  We show  that this tensor  is clearly suited to
  the study of gravitational dynamics and,  moreover, that knowing its
  evolution  is  enough to  completely describe  a fluid  element.  In
  order   to determine   this   evolution,  we  make  some    physical
  approximations on  the exact dynamical system  and we  thus obtain a
  closed  local  system   of differential    equations governing   the
  evolution of the deformation tensor.  Our approximate dynamics treat
  exactly the conservation of mass, of the  velocity divergence and of
  the shear and  is exact  in  the case   of  planar, cylindrical  and
  spherical  collapses.   It  also  reproduces  very  accurately   the
  evolution  of all dynamical  quantities  for a   very wide  class of
  initial conditions as illustrated by  a detailed comparison with the
  homogeneous ellipsoid model.    Beside, we highlight, for the  first
  time, the important  dynamical    role  played by  the     Newtonian
  counterpart of the magnetic part of the Weyl tensor.
\end{abstract}

\section{Introduction}
Two main analytical approaches  have   been used to  study  non-linear
gravitational instability dynamics   in cosmology.  The first   one is
Eulerian, the fundamental quantities are then the density and velocity
fields  evaluated  at the    comoving Eulerian coordinate   ${\bm  x}$
(\cite{Goroff1986},  \cite{Grinstein1987}).  The second method is that
of Lagrangian trajectories (\cite{Zeldovich1970}, \cite{Moutarde1991},
\cite{Buchert1993}, \cite{Bouchet95}).   However, since the  equations
governing the motion of a self-gravitating  dust are highly non-linear
and non-local, finding an exact solution  in the general case, both in
the Eulerian  or Lagrangian  approaches is  impossible.  Consequently,
one  is then forced  either  to  restrict  the problem to   particular
geometries (spherical   or  planar)  or  to  use  approximation
techniques such as a perturbative theory.  In the  first case, one can
study  the very nonlinear evolution of  a single object, but the range
of initial conditions is then  by definition very limited.  The second
case   applies  to any  type of   initial conditions,  it  consists of
expanding the {\it solutions} of dynamical equations.  In the Eulerian
formalism one  develops the density  contrast  and peculiar velocity
fields as follow:
\begin{equation} 
\delta=\sum_{n=0}^{n=\infty} \epsilon_{\delta}^{(n)} \delta^{(n)}
\hskip 2cm
v=\sum_{n=0}^{n=\infty} \epsilon_{v}^{(n)} v^{(n)}
\end{equation} 
These developments are  introduced  in the dynamical  equations (Mass
conservation,  Euler and Poisson   equations).  The equations are then
solved keeping only terms in $\epsilon$  and after in $\epsilon^2$ and
so on.  The  Eulerian expansion of the  density contrast is valid when
$\delta < 1$, but when $\delta \approx 1$, neglecting the term of high
order is no longer possible.  In the Lagrangian formalism the solution
is obtained as a perturbated trajectory about the linear displacement
\begin{equation}
{\bm P ({\bm q},t)} 
= \sum_{n=0}^{n=\infty} \epsilon_{P}^{(n)}{\bm P^{(n)} ({\bm q},t)}
\end{equation}
where {\bm  q} is the  Lagrangian coordinate.   This kind of expansion
allows   one  to follow  the  dynamics  longer   than  in the Eulerian
description, since  the Lagrangian picture is intrinsically non-linear
in  the  density  field.  The  first    order recovers  the  Zeldovich
approximation   (\cite{Zeldovich1970}).   Higher     orders    provide
corrections  but   the calculation     rapidly  becomes  very  tedious
(\cite{Moutarde1991},   \cite{Buchert1992})   and   their  domain   of
application are  in     fact restricted to   the   quasi-linear regime
($\delta$ less than a few).

In this paper we propose a new Lagrangian  approximation. It allows to
compute  the local  properties of a   fluid element during its motion,
eventhough it does  not determine its trajectory.   From this point of
view, our work is in the same spirit as that of Bertschinger $\&$ Jain
(1994)  , however, in our case,  the nonlinear dynamic will be studied
through only one quantity, the deformation tensor
\begin{equation}
{\cal J}_{ij}={\bm 1}_{ij} + {\cal W}_{ij}
=\left(\frac{\partial x_i}{\partial q_j}
\right).
\end{equation}
{\bm x} and {\bm q} are respectively the Eulerian and Lagrangian
coordinates of a fluid  element.  

The deformation tensor   when it is not  singular, contains
all   the dynamical  information   on  a  fluid element excepted   its
trajectory.  From this tensor, and  from its time-derivatives, one can
compute  the density field,  the velocity divergence, the shear tensor
and  the tidal  tensor.   Consequently,  the  influence  of all  these
quantities which  are very operative  during the evolution  of a fluid
element (\cite{BJ1994},   \cite{vdW94}), will then be  naturally taken
into account  in our formalism.   Moreover as  ${\cal  W}_{ij}$ are by
definition always smaller  than $1$ (in absolute  value) ,   for a
collapsing    fluid element,   they   are  very suitable
quantities for all approximate perturbative  description and can allow
one   to follow  the evolution of    large density contrasts or  large
displacements, this     is not possible  with   the  usual Eulerian or
Lagrangian perturbative  approach.     If a fluid     element is
expanding in one direction  the  corresponding ${\cal W}_{ij}$  can be
greater  than  $1$.  In that  respect,  the deformation tensor is more
adapted to describe collapsing regions than expanding voids.

We deduce the nonlinear dynamics of the deformation tensor by using an
alternative  procedure  to the  usual  perturbative  expansion  of the
solutions   of the exact dynamical   system.   Our procedure consists,
after some  physical  approximations, in  deducing directly  from  the
exact system a new approximate  dynamical system where the nonlocality
of the gravitational dynamic  introduced  by the Poisson equation  has
been cancelled.   The new closed  set of local Lagrangian equations is
then solved exactly.  It  takes into  account nonlinear effects  which
could  be absent  in  a  perturbative  solution.   In other  fields  of
research  similar  procedures have  been  used.   It is  the  case for
example, in  the turbulent hydrodynamic or  in the state solid physic,
where some important physical processes, as the generation of solitons
and   the  tunnel  effect respectively,   can  not  be explained  by a
perturbative solution.  Our  system reproduces  exactly the spherical,
cylindrical and planar  collapses, but it   also allows one  to
describe a  much wider class of initial  conditions and to  follow the
dynamics into the highly nonlinear regime.

A comparative  analysis of  our   approximation  with the  results  of
\cite{BJ1994} and Lagrangian  perturbative   solutions has  also  been
performed.  It has  allowed  a better understanding of  the respective
influence of  different  dynamical quantities during  the collapse, as
for example the Newtonian limit of magnetic part  of Weyl tensor which
has been   extensively   discussed in  recent papers   (\cite{BJ1994},
\cite{BH1994}, \cite{KP1995}, \cite{ED1995})

The outline of the paper is the following: In section 2 we present the
dynamical  equations of the deformation  tensor and our approximations
on  these equations.  In section 3  we  give a physical explanation to
our  approximations and  evaluate  their accuracy.  We  also suggest a
classification of collapses according to their geometrical properties.
We compare our results with other  approximate descriptions in section
4 Finally, conclusions  and  a summary  of our  results  are given  in
section 5 The complete perturbative solutions to our dynamics is given
in the appendix.

\section{Dynamics of the Deformation Tensor}
The evolution of the cold matter in an  expanding Universe is governed
by the following set of equations:
\begin{eqnarray} 
& & \frac{\partial \delta}{\partial t} +\frac{1}{a} \nabla_{\bm x}.
(1+\delta){\bm v} = 0 \\
& & \frac{\partial{\bm v}}{\partial t} + \frac{1}{a} ({\bm
  v}.\nabla_{\bm x}){\bm v} + \frac{\dot{a}} {a} \bm{v} = -\frac{1}{a}
  \nabla_{\bm x} \Phi \\ 
& &\nabla_{\bm x}^2 \Phi  =  4 \pi G a^2 \bar{\rho} \delta    
\label{pois1}
\end{eqnarray}
$\Phi$  is the gravitational potential  of the fluctuating part of the
matter density, ($\bar{\rho}$  is the average matter density, $\delta$
is the density  contrast) ${\bm  v}$  is the  peculiar velocity (${\bm
  v}=a\frac{d{\bm x}}{dt})$   and  $a$ is   the  scale  factor.    The
non-locality introduced by   the Poisson equation  (\ref{pois1}) makes
this  system very difficult   to   integrate. However, by  using   the
deformation  tensor  ${\cal  J}$,  it becomes possible   to  derive an
approximate local description.

First, we change  from the Eulerian to  the Lagrangian time derivative
and  we    introduce  the    conformal     time $\tau$   defined    by
$d\tau=\frac{dt}{a^2}$, the previous  set  of  equations can then   be
written as:
\begin{eqnarray} 
\label{cons}
& &\frac{d\delta}{d\tau}+a(1+\delta)\theta=0    \\
\label{eul}
& &\frac{d^2 {\bm x}}{d\tau^2}=-\nabla_{\bm x}\tilde\Phi 
= a\frac{\partial{\bm v}}{\partial \tau}+\dot{a}{\bm v} +
  a^2({\bm v}.\nabla){\bm v} \\
& &\nabla_{\bm x}^2 \tilde\Phi  =  4 \pi G a^4 \bar{\rho} \delta
\label{pois2}
\end{eqnarray} 
where $\tilde\Phi=a^2\Phi$ and  $\theta$  is the velocity divergence.  

The conservation of  mass  (equation (\ref{cons})) can be rewritten
in terms  of ${\cal J}$  as $1+\delta=\frac{1}{J}$  where $J=\det{{\cal
    J}}$.  The  derivative  with   respect to   {\bf q}  of   equation
(\ref{eul})  and the properties of   commutation between the operators
$\nabla_{\bm q}$ and $\frac{d}{d\tau}$ yields
\begin{equation} 
d^2_{\tau}{\cal J}=-{\cal J}*{\cal F}
\end{equation} 
where $*$ denotes a   matrix product.  Finally, by introducing  ${\cal
  F}_{ij}=\nabla_{x_i}\nabla_{x_j}  \tilde\Phi$        in     equation
(\ref{pois2}), the new system of equations becomes:
\begin{equation}
\left\lbrace
\begin{array}{l}
1+\delta=\frac{1}{J}\\
d^2_{\tau}{\cal J}=-{\cal J}*{\cal F} \\
Tr({\cal F})=\beta(\tau)\delta 
\end{array}
\right.
\label{eqc}
\end{equation}
We have defined  $\beta(\tau)=4\pi Ga^4(\tau)\bar\rho(\tau)$.  All the
effects of $\Omega$  on  the dynamics  are contained  in this function
$\beta$.  In an $\Omega=1$ universe  $\beta= \frac{6}{\tau^2}$.  These
equations were  first written  by  Lachièze-Rey  (\cite{MLR1993})  who
proposes a solution where  the deformation tensor has the  particular
form: ${\cal J}({\bm q},t)=D_{\bm q}(t){\cal J}_0  ({\bm q})$.  ${\cal
  J}$  is  a tensor  field   which  depends   only on  the  Lagrangian
coordinates and $D_{\bm q}(t)$ is  a scalar function  of time which is
given  by a differential equation  whose coefficients are functions of
the eigen-values   of ${\cal  J}_0({\bm  q})$  .   Assuming   that the
deformation tensor remains proportional   to its initial value  during
the evolution  of the fluid element greatly  simplifies  the dynamics. 
However,  this is  a very  restrictive assumption,  and such a solution
admits  a perturbative expansion which generally  differs even at second
order from the usual Lagrangian perturbative solution.

In  the general case  ${\cal J}$  has  nine independent components and
system   (\ref{eqc})  cannot be  solved.   However,   if  the  flow  is
irrotational   then  ${\cal S}_{ij}   = \frac{1}{a}\frac{\partial{\bm
    v_i}}{\partial{\bm  x_j}}$   is   symmetric and   since  ${\cal   S}
=\frac{1}{a} {\cal J}^{-1}*d_{\tau}{\cal J}$, this implies that ${\cal
  J}$ is also symmetric, with only six independent components.  In the
rest of  this  paper we    restrict ourself  to  irrotational flow.    
Moreover, if  ${\cal  J}$ is  symmetric, it can  be  diagonalized.  We
perform this diagonalization in the initial conditions and we consider
that  the  first  and  second  order  time-derivative  of non-diagonal
components are  initially vanishing, in order  that ${\cal J}$ remains
diagonal during all the evolution.  This assumption restricts somewhat
the range of  initial  conditions.  However, as it  has  been shown by
Moutarde  {\it et al.}  (1991) in the case  of the  evolution of three
caustics the non diagonal terms  which appear during the evolution, do
not modify strongly the dynamics of the diagonal terms.

From a  diagonal deformation tensor ${\cal J}_i  = 1_i + w_i$  all the
usual dynamical quantities can then be easily expressed 
\begin{equation}
\delta = \frac{1}{J} - 1 = \frac{1}{(1+w_1)(1 + w_2)(1+w_3)} -1 
\end{equation}
\begin{equation}
\theta=\sum_{i=1}^{3} \frac{\dot{w_i}}{a(1+w_i)} 
\end{equation}
\begin{equation}
\sigma_{i}=\sigma_{ii}=\frac{\dot{w_i}}{a(1+w_i)}-\frac{1}{3}\theta
\end{equation}
\begin{equation}
{\cal F}_{i}={\cal F}_{ii}=-\frac{\ddot{w_i}}{(1+w_i)}
\end{equation}

where $\theta$ and  $\sigma$ are respectively  the velocity divergence
and the shear tensor.  It  is important to  note that with the above
expressions a  major  part of  the  dynamics is then  already  exactly
described.     The   conservation-evolution   equations   of  $\delta$
(continuity equation), of   $\theta$, (Raychaudhuri  equation)  and of
$\sigma$, are in  fact automatically satisfied.  We  only have to find
the  evolution  equations for the $w_i$  components.  Such an equation
will be  derived from an local approximation  of the non-local Poisson
equation.

Equations (\ref{eqc}) give:
\begin{equation}
Tr({\cal F})=-Tr({\cal J}^{-1}d^2_{\tau} {\cal J})
=\beta({\tau})(\frac{1}{J}-1)
\end{equation}
In terms of $w_i$ this equation can be written:
\begin{eqnarray} 
& &\sum_{i=1,3} \lbrace (1 + w_j + w_k + w_jw_k) d^2_{\tau}w_i \nonumber \\  
& & \hspace{1.6cm} = \beta(\tau) (1+\frac{w_j+w_k}{2} 
                         + \frac{w_jw_k}{3})w_i
\rbrace
\label{eqd}
\end{eqnarray} 
where $(i,j,k)$ is a circular  permutation of $(1,2,3)$. This equation
does  not contain enough  information    in  order to evaluate    the
evolution  of   each  $w_i$   component.  However  its   re-writting,
respectively,    in  one-dimensional,  two-dimensional      and  three
dimensional geometries   suggest  reasonable physical  approximations,
from which we will deduce an  approximate evolution of the deformation
tensor.

In a one-dimensional geometry, equation (\ref{eqd}) reduces to:
\begin{equation} 
d^2_\tau w_1=\beta(\tau)w_1\Rightarrow w_1(q,t)=a(t)w_1^0(q)
\label{eq1d}
\end{equation}
where $w_1^0(q)$   is  the   Lagrangian  deformation   field  linearly
extrapolated to $a=1$.  This  solution  is exact for a  planar
collapse.   A similar evolution  for the three  components of $\cal W$
tensor corresponds to the Zeldovich approximation.

In a two-dimensional geometry, equation (\ref{eqd}) can be written, where
$(i,j)$ is a circular permutation of $(1,2)$ 
\begin{equation} 
\sum_{i=1,2}\lbrace
(1+ w_j ) d^2_{\tau}w_i  = \beta(\tau) (1+\frac{w_j}{2})w_i
            \rbrace
\label{eqe}
\end{equation} 
In this equation   $w_1$ and $w_2$   have a symmetric  role. Splitting
symmetrically this unique equation in two, we get:
\begin{equation}
(1+w_j)d^2_\tau w_i=\beta(\tau)(w_i+w_iw_j)+B_{ij}
\label{eq2d} 
\end{equation} 
In these equations $B_{ij}$ components are unknown, however, they must
obey the following constraints:
\begin{enumerate}
\item $B_{ij}=-B_{ji}$ because the sum  of equations (\ref{eq2d})  must
  reduce to equation (\ref{eqd}).
\item $w_j=0$ implies $Bij=0$ in  order to be  consistent with the one
  dimensional case.
\item $w_i=w_j$  implies $B_{ij}=0$.  In  this  case we have  only one
  unknown and equation (\ref{eqd}) gives the solution.
\end{enumerate}           
   
Making the hypothesis that $B_{ij}$ can be expressed as a polynomial of
$w_i$ the previous constraints gives:
\begin{itemize}
        \item 3. $\Rightarrow B_{ij} \propto (w_i-w_j)$
        \item 2. $\Rightarrow B_{ij} \propto w_j(w_i-w_j)$ 
        \item 1. $\Rightarrow B_{ij} \propto w_iw_j(w_i-w_j)$
\end{itemize}

This means that $B_{ij}$ is at least of third order in $w_i$. Equation
(\ref{eq2d}) with  $B_{ij} =  0$  is  then  exact  up to second  order
included   in $w_i$ terms.   It  can therefore be   used to compute the
evolution of the deformation tensor with a very good accuracy.

Following the same procedure in three dimensional case we get the three
following equations $(i=1,2,3)$:
\begin{eqnarray} 
& &(1 + w_j + w_k + w_jw_k) d^2_{\tau}w_i \nonumber \\  
& & \hspace{1.6cm} = \beta(\tau) (1+\frac{w_j+w_k}{2} 
                         + \frac{w_jw_k}{3})w_i + B_{ijk}
\label{eqf}
\end{eqnarray} 

This time the constraints on $B_{ijk}$ are:
\begin{enumerate}
\item $B_{ijk}=B_{ikj}$,  because $w_j$ and  $w_k$ play symmetrical roles
  towards $w_i$.
\item $w_j=w_k=0$ implies $B_{ijk}=0$  in order to be  consistent with
  the one dimensional case.
\item $w_i=w_j=w_k$ implies $B_{ijk}=0$,  because  in this case,  which
  corresponds to a spherical geometry,  there  is only one unknown  and
  equation (\ref{eqd}) then gives the exact solution.
\item $w_k=0$ implies $B_{ijk}=B_{ij}$ in order  to be consistent with
  the two dimensional case.
\item  $B_{ijk}+B_{jki}+B_{kij}=0$,  because the sum of  the equations
  (\ref{eqf}) must reduce to equation (\ref{eqd}).
\end{enumerate}   

Making again the  hypothesis that $B_{ijk}$  is a polynomial in terms of
$w_i$, the polynomial of lowest  order satisfying the constraints 1.) to
5.) is $B_{ijk}=B_{ij}+B_{ik}$.

Equations (\ref{eqf}) with $B_{ijk} = 0$ 
\begin{eqnarray}
& &(1 + w_j + w_k + w_jw_k) d^2_{\tau}w_i \nonumber \\  
& & \hspace{1.6cm} = \beta(\tau) (1+\frac{w_j+w_k}{2} 
                         + \frac{w_jw_k}{3})w_i
\label{eq3d}
\end{eqnarray}
are then exact at  least up to second  order included in $w_i$ terms.  
They  form a close system of  perturbed equations.  It is important to
notice  that our approximation is  both perturbative  and geometrical. 
It can be considered  as perturbative because we  discard a term of at
least third order in $w_i$ in equation (\ref{eqf})  and it can also be
considered   as a geometrical approximation   because  it is exact for
planar, cylindrical and spherical collapse and only approximate
in the  general  case.  In the  next section  we discuss  the physical
meaning  of the approximate  dynamics of equations (\ref{eq3d}) and we
show   that  our solution reproduces    quite faithfully the nonlinear
dynamic, in particular the evolution of density contrast.

\begin{figure*}
\epsfxsize=18.0cm
\epsfbox[33 38 576 361]{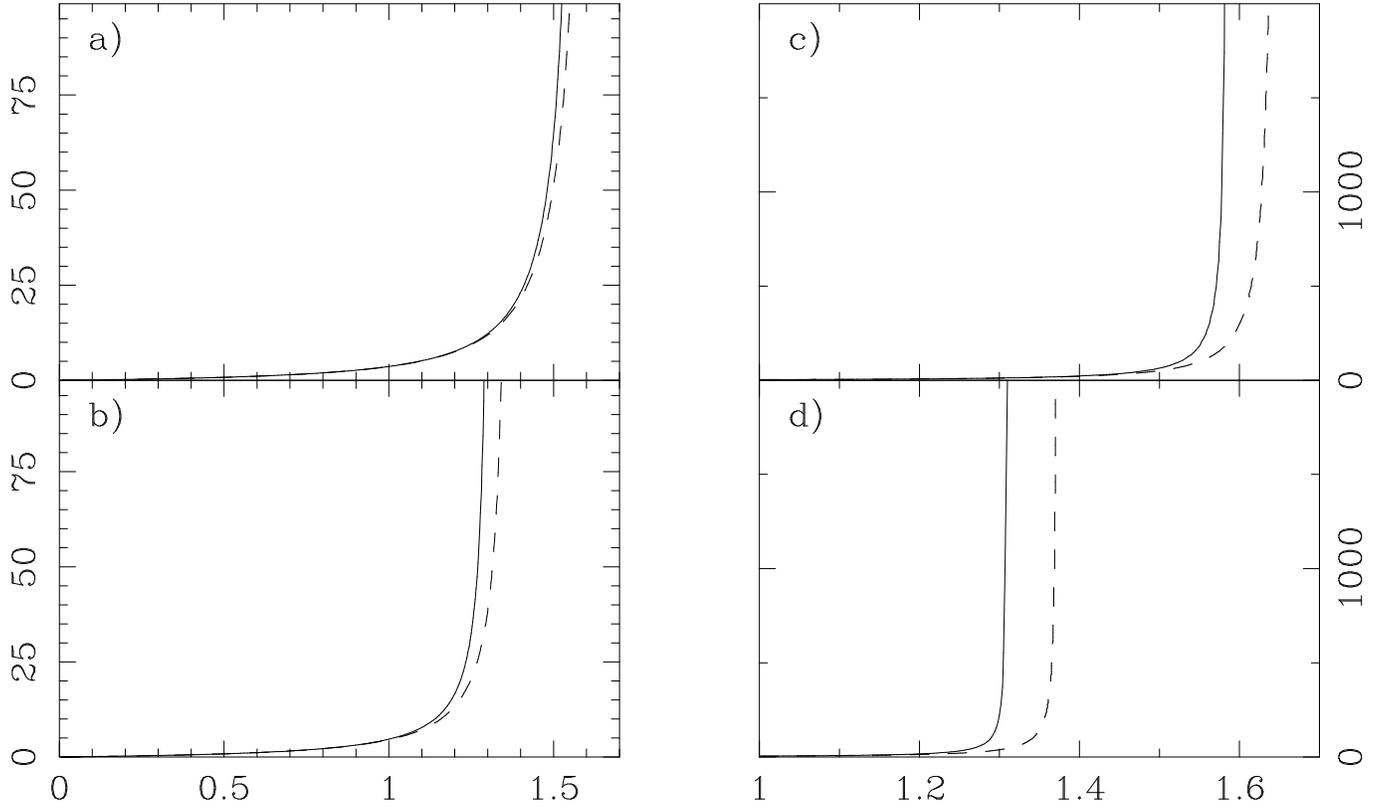}
\label{evd}
{\bf
\caption{This figure shows the evolution of the density contrast for two
  different ellipsoids.  The y-axis is labeled in terms of the density
  contrast and the  x-axis in terms of $\tilde{a}=a*\delta_i/a_i$ where
  a  is  the scale factor  and  $\delta_i$ and $a_i$   are the initial
  density    contrast  and  scale  factor.   The  axis   ratios of the
  ellipsoids  are 1.24-0.8 for a) and c) and 2.21-0.45  for 
  b), and d)} }
\end{figure*}

\section{Physical interpretation of the approximation}
\subsection{Spherical and ellipsoidal collapses}
The    equations   (\ref{eq3d})   which     reduces   respectively  in
one-dimensional   and   two-dimensional    geometries   to   equations
(\ref{eq1d}) and (\ref{eq2d})  (with $B_{ij}=0$), reproduces the exact
planar   and  cylindrical dynamics.   In  spherical   geometry,
equations (\ref{eq3d}) also gives the exact solution.   As a matter of
fact, in such a case,  there is only  one free parameter, the  density
contrast $\delta$ which is directly related to the unique component of
the deformation  tensor, $w_{spherical} = (\frac{1}{1+\delta})^{1/3} -
1$.   Such a relation  allows one then   to identify  the  collapse of  a
homogeneous sphere with a density contrast $\delta$  and a collapse in
one point with a   spherical  environment defined  by  $w_{spherical}$
whose evolution is given by equations (\ref{eq3d}).  In the following,
in a  similar way, we will identify  finite scale collapses with local
collapses with the same geometry.

When the three  diagonal components of  $\cal W$ are  different, it is
natural to  generalize  the  sphere to  an ellipsoid.   The  equations
(\ref{eq3d}) then describe the collapse  of  a point with an  elliptic
environment  which means that the deformation   tensor at such a point
can be   identified with the deformation  tensor  at the center  of an
homogeneous ellipsoid.  The question is now to find such an ellipsoid.
An ellipsoid  of finite size is  defined by three free  parameters as,
for example, the two axis   ratios, and the initial density   contrast
(the local dynamical quantities at the center  of the ellipsoid do not
depend on the size  of the ellipsoid  but  only on its  shape).  These
three  free parameters are not directly  related to  the initial eigen
values of the deformation tensor.   Consequently in order to link  the
ellipsoid of finite size and   the initial deformation tensor we  have
used the  tide tensor.  This one is  very meaningful for the dynamics
and it is well defined in  both cases.  The  tide tensor at the center
of the ellipsoid can  be computed using  elliptic integrals, and it is
related to  ${\cal W}$   by $  {\cal  F}  = -(  {\bm   1} + {\cal  W})
d^2_{\tau} {\cal W}$.  In Figures  1 we compare  the evolution of  the
density contrast computed    from  equations (\ref{eq3d}) and   for  a
homogeneous  ellipsoid having  the  same   initial tide  tensor.    In
principle the   ellipsoid  generates inhomogeneities   in the  outside
homogeneous background and these inhomogeneities in return perturb the
density   inside  the  ellipsoid.    The  dynamics of  the homogeneous
ellipsoid is computed by neglecting such a feedback effect which means
that the density inside and outside the ellipsoid remains homogeneous.
A  detailed analysis    of the ellipsoidal     model can be  found  in
\cite{I73}, \cite{WS79}.   In Figure 1,  we see that our approximation
reproduces very  accurately the evolution of   the density contrast at
the beginning  of the  non-linear  regime even  for very  non spherical
cases  (Figure 1b).  The   error   is always   of  a few  percent  for
$\delta=10$.  In the very non-linear phase  (for $\delta \approx 100$)
the dynamical  delay of our approximation  is, in all cases, less than
5\%.  Figures 1c-1d show the final stage of the collapse. The collapse
time  when the density diverges, is  estimated with an accuracy better
than 5\%. This one remains rather constant, whatever the ellipticity.

\begin{figure*}
\leavevmode
\hbox{%
\epsfbox{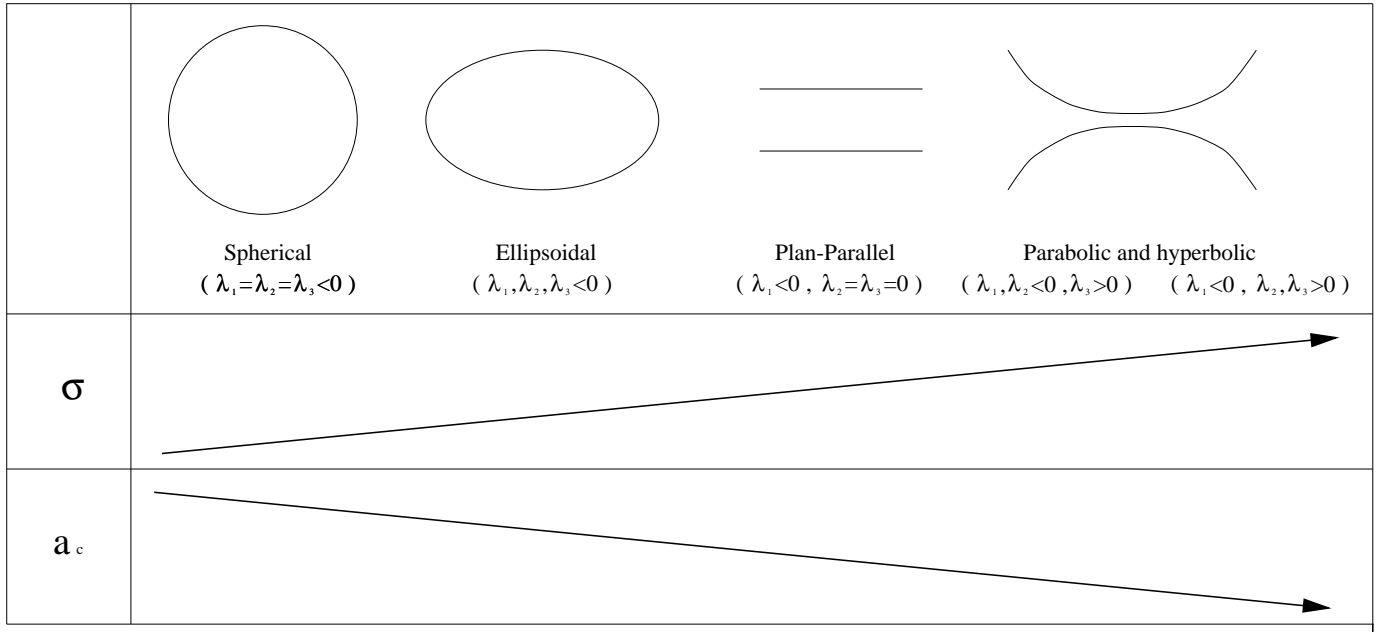}}
\label{table1}
{\bf
\caption{This figure shows the geometrical classification of the
  different possible collapses. $\sigma$ represents the eigenvalue of
  the shear on the fastest collapsing direction and $a_c$ the collapse
  time of the fluid element. (The initial density contrast is supposed
  to be positive and equal in all cases).}
}
\end{figure*}

This  generalization of the spherical collapse  allows one to describe
points   with an anisotropic   deformation  tensor.  The matter
distribution around   the   fluid  elements  is  approximated    by an
ellispsoid  which can  fit  a generic matter  distribution much better
than   a  sphere. The picture  of an     ellipsoid is helpful  in
understanding  the dynamics  but one   should not forget  that we  are
evolving Lagrangian  fluid elements and  not a finite sized ellipsoid. 
Another important  point is that  for an  elliptic  collapse the  tide
tensor remains diagonal, this  was  our first approximation to  obtain
equations (\ref{eq3d}). Consequently, if  we restrict ourself to  this
kind of   collapse our only approximation  is  $B_{ijk}=0$ and we have
shown that discarding this term was a very good approximation.

\subsection{Others Geometries}

The evolution described by equations (\ref{eq3d}) is not restricted to
the Lagrangian fluid elements with an elliptical environment for which
the three diagonal components  of $\cal W$ are  of the same sign (e.g. 
the opposite sign of $\delta$). It can be applied to  a wider class of
initial conditions   corresponding to an     initial ${\cal W}$   with
eigenvalues of different sign.  We call  parabolic (resp.  hyperbolic)
the points which have an  initial deformation tensor with two negative
(resp.  positive) and  one positive (resp.  negative) eigenvalues. For
these points which  correspond to  more complex  geometries it is  not
possible to find a simple distribution of matter with which they could
be compared. They might be  viewed as ellipsoids  subject to a  strong
external  tidal  fields caused by  surrounding   matter.  If the tidal
field is  strong  enough  to make  one  direction   expand instead  of
collapsing the  global matter distribution  is not well describe by an
ellipsoid.  The parabolic and hyperbolic geometries allows to describe
such cases.

The transition from elliptic to parabolic collapse is the planar ({\it
  i.e.} Zeldovich) collapse.  As illustrated in Figure 2, the shear on
the fastest   collapsing  direction increases and  the   collapse time
decreases  when one  goes  from elliptic to   parabolic and finally to
hyperbolic  points.  Parabolic and  hyperbolic points collapse even if
they are  initially underdense due  to the effect  of the shear and of
the tide.   This illustrates that  these two quantities,  and not only
the density  contrast, play  a key  role  in the dynamics.   The  only
points  which do not reach  infinite density are elliptic points which
are initially  underdense  or equivalently points with  three positive
eigenvalues for their  initial deformation tensor.  As in the case
of  the Zeldovich   approximation, all  the  points  with one  initial
negative   eigenvalue for their  deformation  tensor, collapse.  These
points represent 92\% of  all points.  This  ratio can be  computed by
integrating    the distribution  function   of  the deformation tensor
eigenvalues obtained  by Doroshkevich (\cite{doro}). Excepting
cylindrical and spherical  points  which are highly   symmetrical, the
direction   with the  most negative   initial  eigenvalues of $\cal W$
collapses first  and  alone.  Consequently the collapsed  objects will
have a pancake shape.

In the  next section we concentrate  on elliptic  collapses and make a
detailed  comparison  of different approximations   using not only the
density contrast but also the shear and the tide tensor.

 
\section{Comparison of different approximations:}

In the previous section we have seen that our approximation reproduces
with a  good  accuracy the  evolution  of the density contrast  at the
center of a homogeneous ellipsoid (for fluid elements with an elliptic
geometry).  We will now push further this  analysis in two directions. 
First,  we will compare our   approximation (hereafter VB as Vanishing
$B_{ijk}$), the one proposed  by Bertschinger and Jain (\cite{BJ1994})
(hereafter VH as Vanishing  ${\cal{H}}_{ij}$) and the first and second
order developments  (these developments  are  obtained by  the  method
presented in the Appendix) to  the homogeneous ellipsoid model.  Since
all these approximations keep the  deformation tensor diagonal, it  is
natural to  compare   them  with  the  dynamics of    the  homogeneous
ellipsoid.  Secondly, we will show that the density contrast alone can
be  misleading  in determining  the  exactitude of   a given dynamic.  
Consequently we  will also study the  evolution of the shear  and tide
tensor  which, as we have seen,  have  a very important dynamical role
and give valuable information about the geometry of the collapse.

To make  the  discussion  more   explicit   we describe  briefly   the
approximation suggested by Bertschinger  and Jain. These  authors have
also proposed  a closed local set  of Lagrangian equations to describe
the  evolution of a  fluid element.   This   set of equations  for the
irrotationnal  case  and for  diagonal shear  and  tide tensor  can be
written as:

\begin{eqnarray}
\label{b1}
& & \dot{\delta}+a(1+\delta)\theta=0 \\   
\label{b2}
& & \dot{\theta} + \frac{\dot{a}}{a}\theta + \frac{a}{3}\theta^2 + 
a\sigma^{i}\sigma_{i} =-4\pi G \bar{\rho} a^3 \delta \\ 
\label{b3}
& & {\dot{\sigma}}_{i} + \frac{\dot{a}}{a}\sigma_{i} + \frac{2a}{3} 
\theta\sigma_{i} + a \sigma_{i}^{2} 
-\frac{a}{3}(\sigma^{k}\sigma_{k}) = -a {\cal E}_{i} \\
\label{b4} 
& & \dot{{\cal E}_{i}} - \frac{\dot{a}}{a}\cal E_{i} + a \theta {\cal E}_{i}
+a \sigma^{k} {\cal E}_{k} -3 a \sigma_{i} {\cal E}_{i} = -4 \pi G \rho a^5
\sigma_{i} 
\end{eqnarray} 

where ${\cal E}_{ij}$ is the traceless part of  the tide tensor ${\cal
  F}_{ij}=\nabla_{x_i}\nabla_{x_j}   \tilde\Phi$ (${\cal E}_{ij}={\cal
  F}_{ij}-\frac{\delta_{ij}}{3}\nabla^2   \tilde\Phi$)   and the   dot
denote  derivation    with      respect   to    $\tau$.      Equations
(\ref{b1})-(\ref{b3})           are     easily     deduced        from
(\ref{cons})-(\ref{pois2}).   Equation   (\ref{b4})     governing  the
evolution of the tide tensor is  obtained in the Newtonian limit, from
general relativity  by neglecting the magnetic part (${\cal{H}}_{ij}$)
of the Weyl tensor (we will hereafter call  $M_{i}$ the term discarded
in order to obtain equation (\ref{b4}) divided  by the sum in absolute
value of all the terms of this equation).  The Newtonian expression of
the Weyl tensor has been abundantly discussed recently (\cite{KP1995},
\cite{BH1994}) but its  dynamical  influence remains  unclear.  During
our analysis we will give a partial answer to this question.

\begin{figure*}
\epsfxsize=18.0cm
\epsfbox[29 48 582 624]{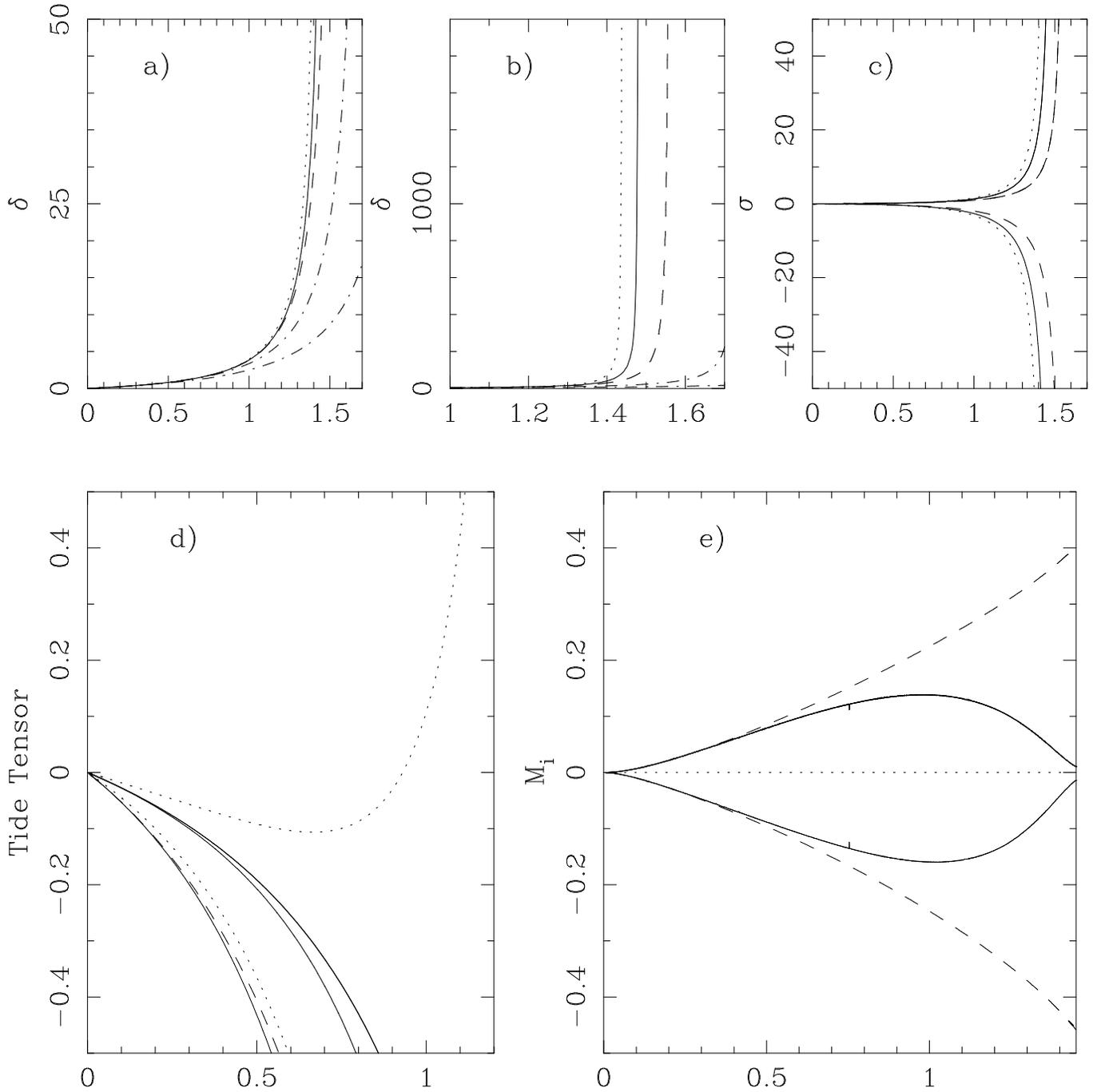}
\label{dat3}
{\bf
\caption{This figure displays the evolution of different dynamical
  quantities  for the homogeneous  ellipsoid model (full line), for VB
  (dashed line),  for VH (doted  line) and for  first and second order
  perturbative  solutions  (dashed-doted  line). The  x-axis is always
  labeled in  terms of $\tilde{a}$  (cf. fig \ref{evd}). Graph a) and
  b)  shows  the  evolution of  the   density contrast  with different
  scales, graph c)  and d) shows  the evolution of the  eigenvalues of
  the shear and tide  tensor and  graph  e) the evolution of  $M_i$ as
  defined in the  text. The initial axis ratio  of the ellipsoid  are:
  1-0.58. } }
\end{figure*}

\begin{figure*}
\epsfxsize=18.0cm
\epsfbox[29 48 582 624]{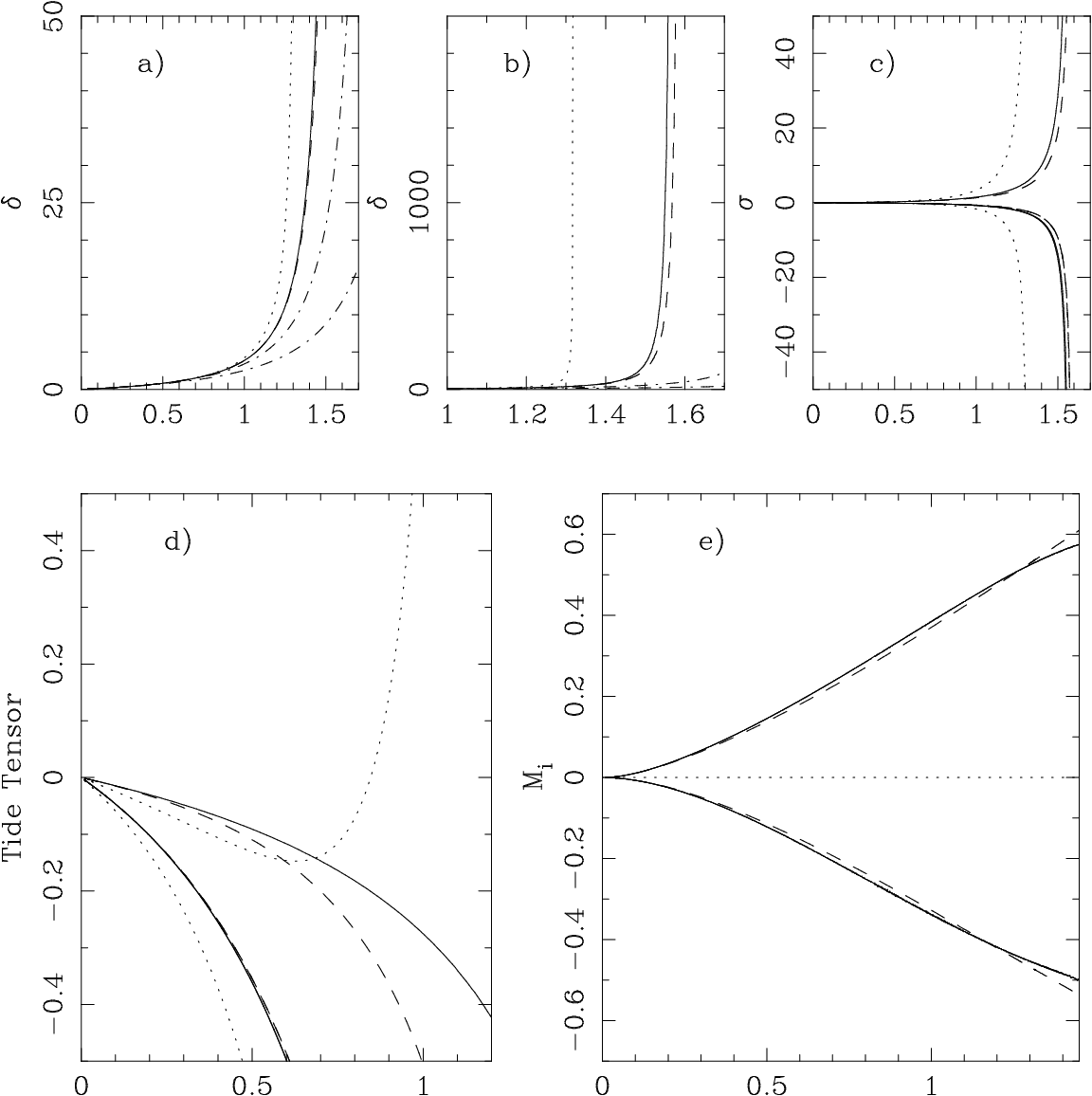}
\label{dat7}
{\bf
\caption{This figure displays the same graphs as in figure 3,
  but for an ellipsoid with an initial axis ratio: 2.4-1}
}
\end{figure*}

Let us first examine the case of an ellipsoid which collapses first in
one direction, the two other  directions being equal. The ellipsoid is
like a  thick disc which  collapse along  its  axis of  symmetry.  The
evolution of  the density contrast is perfectly   described both by VB
and  VH even in the non-linear  phase ($\delta \approx 50  $) (Fig.  3
a)).  The improvement compared  to first and second order developments
is significant, which illustrates that integrating perturbed equations
is more  accurate than   finding   a perturbative  solution.    At the
collapse both VH and VB have a dynamical delay of a few percent (Fig 3
b)).    VH under-estimates the   collapse   time and is  slightly more
accurate than VB   which on the  contrary over-estimates  the collapse
time.  The analysis of the shear components gives similar conclusions.
The shear is  well described at the  beginning of the non-linear phase
and has a small dynamical delay at collapse (Fig  3 c)).  However, the
evolution of  the tide tensor gives   quite a different point  of view
(Fig 3 d)).  First,  we see that VH  does not reproduce  correctly the
evolution of  ${\cal{E}}_{ij}$  even at a  very  early stage.   In the
linear  regime   $\Phi   \propto  a^{2}$  which  implies   $\dot{{\cal
    E}}_ij=0$.   Thus,  the first non-vanishing   terms in $\dot{{\cal
    E}}_ij$ are at least  of second order.  Consequently the  dynamics
of  ${\cal  E}_ij$ is much     affected by ${\cal{H}}_{ij}$  which  is
generally not vanishing at second order.  On the other hand, in VB the
term which is  neglected  is at  least of  third order it  is therefor
normal that the  correct evolution for  $\cal{E}_{ij}$ is recovered at
early stage.  In the quasi-linear  regime VB still  gives quite a good
description for  the evolution of the tide   tensor while VH  does not
reproduce the correct behavior.  All these  features can be understood
by looking at   the evolution of  $M_i$  (Fig 3 e))  $M_i$  grows very
rapidly in  the linear regime  which translates immediately in a wrong
evolution equation for the tide  tensor for VH.   VB gives the correct
evolution of $M_i$ in the linear phase.   Near the collapse, since one
direction collapses first, the  fluid element evolves towards a planar
geometry.  For this reason $M_i$ is again vanishing (compared to other
quantities) at the collapse.  VB is unable to follow this behavior and
largely over-estimates the amplitude of $M_i$ at the collapse.

\begin{figure*}
\epsfxsize=18.0cm
\epsfbox[29 48 582 624]{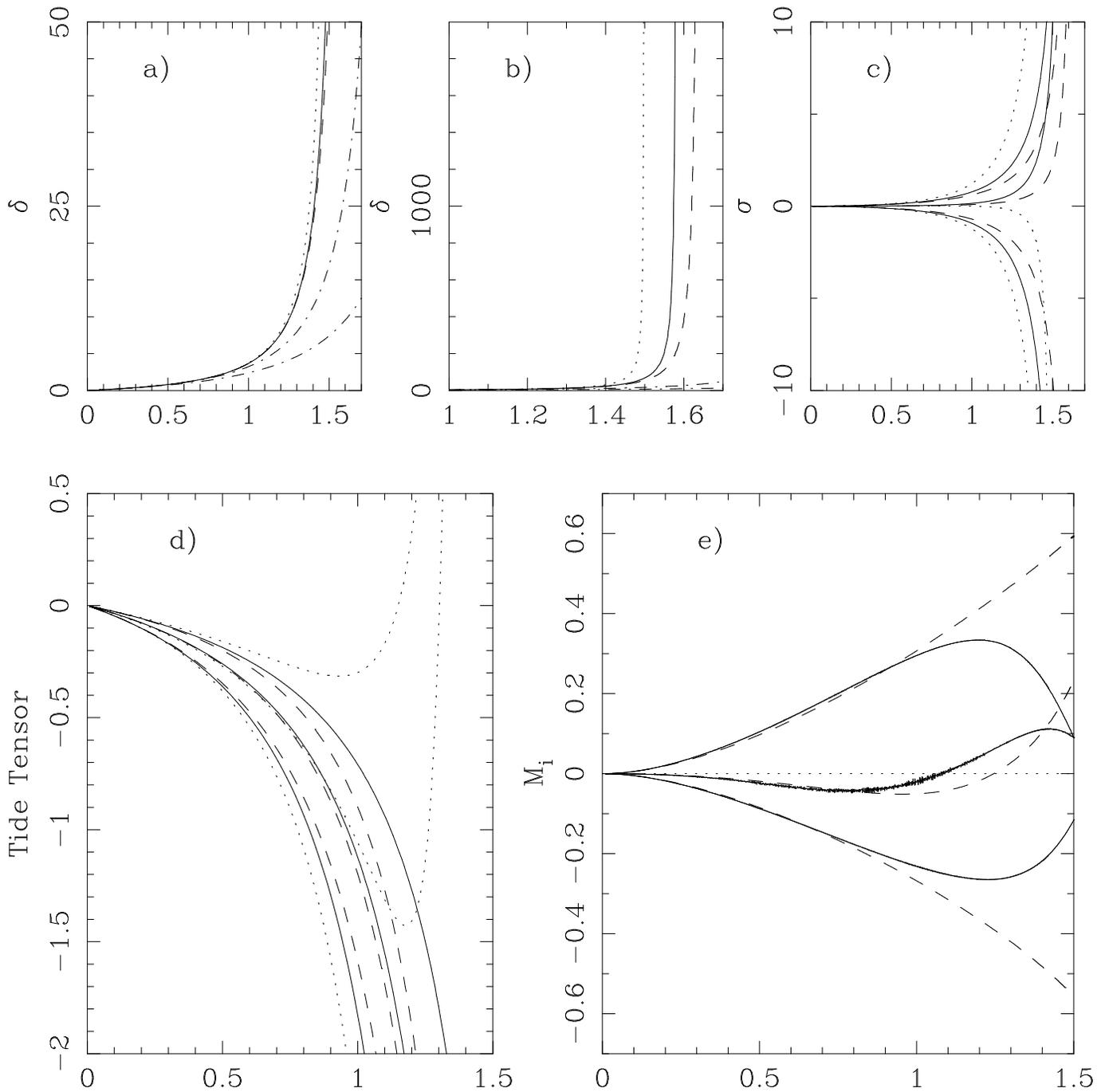}
\label{dat9}
{\bf
\caption{This figure displays the same graphs as in figure 3,
but for an ellipsoid with an initial axis ratio: 1.24-0.8 }
}
\end{figure*}

Now we will  turn our attention to the  case where  two directions are
equal and  collapse first.  The  ellipsoid then has  a cigar shape and
collapses in the two directions perpendicular  to its axis of symetry. 
For  this  particular geometry  VB  is extremely  close   to the exact
(ellipsoidal) dynamics.  The density contrast and the shear tensor are
computed to a  high accuracy (Fig 4  a),  b) and c)).  For  both these
quantities the dynamical delay at the collapse is less than 1\%. VB is
excellent in this case because the initial ellipsoid has a cigar shape
and its dynamics  is close to that  of a cylinder (VB  is exact for  a
cylindrical collapse).  Looking now at the tide tensor we see that the
two directions which collapse first are  perfectly described while the
third  one, which  should be  zero for  a  cylinder, is  not very well
followed.  VH does not give accurate results  in this case.  The fluid
element  tends, at collapse, towards  a cylindrical geometry for which
the   Weyl tensor  is  not   vanishing  (contrary   to  what  said  by
Bertschinger and  Jain in  their  abstract).  Consequently,  $M_i$  is
playing an  increasing  dynamical role  from the linear  phase to  the
collapse which makes  equation (\ref{b4}) a  bad approximation.  VB is
extremly accurate for  this particular ellipsoid  because its dynamics
is very close to  that of a cynlindar and  we  have shown that VB  was
exact in cylindrical  geometry.

The two cases that we have studied previously were highly symmetrical.
They were instructive in order to understand the role of the different
dynamical quantities in each of the  approximations. However, they are
not  representative     of  generic  cosmological   fluid    elements. 
Consequently,  we  will now  study  the  collapse of   an ``ordinary''
ellipsoid with three different axis.  We have plotted  in figure 5 the
evolution of only one  such ellipsoid but  the description we  give is
general and is  the result of the   analysis of many  different cases. 
For these geometries  both  VH and VB  describe  the evolution of  the
density field very accurately (Fig  5 a) and  b)).  VB is a little more
accurate  in the  non-linear  phase and  significantly better for  the
estimation of  the collapse  time.   The error  on  the collapse  time
obtained from VB is always less than 5\%, while it can be much greater
for  VH.   VB always   overestimates  the collapse   time while it  is
underestimated  by VH.  The improvement  compared  to the second order
perturbative solution is in  all  cases very  important.  This  can be
explained by    the fact  that  a  perturbative   solution of equation
(\ref{eqf})     only   preserves   the  perturbative  aspect    of our
approximation but loses its geometrical properties which are ``coded''
in the   equations.   A perturbative   solution  is  inexact  for  the
cylindrical and  the    spherical  collapse.   The    quality  of  the
description of the shear and of the tide given by  VB is quite similar
to that   of the density  contrast  (Fig 5c) and   5d))  .  Both these
quantities  are very well followed  at the beginning of the non-linear
phase and have a small dynamical delay at the  collapse (this delay is
a  little   greater for  the    tide).  Eventhough  our  approximation
reproduces quite  faithfully the dynamical evolution  of all the usual
dynamical quantities, it is  less accurate for   $M_i$.  In the  first
phase of the collapse  the three components  $M_i$ start to grow, this
phase  is very well  followed by VB.   However, in a second phase, the
$M_i$  components return     to   zero under  the   correct   dynamics
(ellipsoidal model) while they continue to grow in VB (Fig  5 e)).  As
was mentioned before, when one direction collapses first the ellipsoid
approaches  a   pancake shape  when  reaching   the collapse and  this
explains why $M_i$ evolve towards zero. However, even  if they tend to
zero at  the collapse the $M_i$  terms have a very important dynamical
role.   Eventhough $M_i$  terms are small  (but not  vanishing) in the
linear  regime,   they increase very   rapidly to   become  one of the
dominant terms in the  equations (\ref{b4}).  Consequently, VH rapidly
becomes a  very bad approximation.   This explains  the rather strange
behavior of the tide  tensor predicted by  VH (Fig 5 d)).   The error
initially  present   in  equation (\ref{b4})  propagates   to equation
(\ref{b3}) and  then to  (\ref{b2}) and  finally to   (\ref{b1}).  One
component of the shear  is much affected by  this error (Fig 5 c)), VH
predicts  two negative components for  the  shear while there is  only
one. The density contrast is less sensitive to it.

From the analysis of the different approximations we can conclude that
the $M_i$ terms slow down the  collapse and couple  the directions.  VB
which overestimates the amplitude of  $M_i$ always collapses too slowly
and the  three components of its tide  tensor are much closer  than in
the exact ellipsoidal dynamics (Fig 5  d)).  On the contrary, VH which
completely discards  $M_i$, collapses too  quickly. Figures  5  c) and d)
illustrate  the  importance of   $M_i$ for the  dynamics and   for the
geometry of the collapsed objects.

\section{Conclusion} 

In this paper  we  have presented  and  tested a new  local Lagrangian
approximation   to  the dynamics  of   a self-gravitating cosmological
fluid.    One originality  of   our  approach   was to  describe   the
gravitational   dynamics through  only  one  quantity, the deformation
tensor. We have shown that  with this tensor it  is possible to write
the dynamical equations in a synthetic form and to highlight the role of
different important dynamical quantities, especially the shear and the
tide. The evolution of  the  deformation is  evaluated by  deriving a
close set of local equations from the exact dynamical system after two
approximations.  The first  one  supposes that the deformation  tensor
remains diagonal   during its  evolution, the   second one consists  in
discarding terms of  third order in  terms  of the   deformation tensor
component.  The {\bf  \it  approximate dynamics} thus   obtained, which
reproduce exactly    the planar,   cylindrical   and spherical
collapses is then exactly   integrated. Such a procedure is   completely
different from  the method which    consists  in deriving  a {\bf   \it
  perturbative solution} from the exact dynamics.

In order  to  test our  approximation   we have compared  it  with the
collapse  of  a  homogeneous  ellipsoid   and  with the  approximation
proposed by Bertschinger \& Jain.  From this analysis we conclude that
our approximate dynamics is a very accurate tool for following all the
dynamical quantities during the collapse.  We have also, for the first
time, shed some light  on the important  dynamical role played by  the
Newtonian  counterpart  of  the  magnetic part   of the  Weyl  tensor,
discarded by    Bertschinger  \& Jain.  This   tensor   slows down the
collapse  and  couples   the directions,  discarding   it reduces  the
collapse  time significantly and greatly  modifies the geometry of the
collapse.  The main weakness of  our approximation is that, unlike the
Zeldovich approximation, it is unable to reconstruct the trajectories.
However  many cosmologically   interesting  quantities,  such  as  for
example the mass function,  can be computed without reconstructing the
final Eulerian  field.  Furthermore one  of the main interests of this
work might  is that  it may give   a better understanding  of the very
complex  phenomena which   occur  in  the   non-linear  phase of   the
gravitational dynamics.

We would like to thanks M. Maccormick for rereading the manuscript.

\section{Appendix: Link with perturbative theory.}

In   section $4$  we  have seen  that integrating  our  system is more
accurate  than  finding a perturbative   solution, at first and second
order, of equations (\ref{eq3d}).  Would this still be true for higher
order solutions and does the  perturbative solution of $n^{th}$  order
converge  toward  the  exactly     integrated solution  of    equation
(\ref{eq3d})?  In order to answer these questions we have developed an
analytical recursive formula which relates the perturbative solution of
equation (\ref{eq3d}) at order $n$  to the solution  at order $(n-1)$. 
The   convergence of  this  development and    the  ability of such  a
perturbative  solution to  reproduce all  the  features of the exactly
integrated solution will be the object of a forthcoming paper.

In this section we give a recursive formula to compute the solution of
our system at any order.  Of course this  development will differ from
the  development of the  exact  dynamics beyond  second order, but  by
going to higher orders we can analytically recover the accuracy of the
numerically integrated solution.

Equations (\ref{eqf}) can be written, after some simple algebra, as:
\begin{eqnarray} 
\label{eqmod} 
& &(1+w_j+w_k+w_jw_k)(d^2_\tau w_i- \beta(\tau)w_i) \nonumber \\
& &\hspace{2.3cm}
=-\beta(\tau) (\frac{w_j+w_k}{2} + \frac{2}{3}w_jw_k)w_i
\end{eqnarray} 
It is possible to  find a perturbative solution {\bf  at any order} of
the preceding   equation.    Let us first   introduce  a  few useful
notations:
$$
w_i=\sum_{n=1}^{\infty}w_i^{(n)}
$$
$$
\alpha_{ij}^{(n)}=\sum_{l=1}^{n-1}w_i^{(l)}w_j^{(n-l)
}
$$
$$
\beta^{(n)}=\sum_{m=1}^{n-2} \sum_{l=1}^{n-m-1} w_i^{(m)}w_j^{(l)}
w_i^{(n-l-m)} 
$$
$$
P_i^{(n)}=d^2_\tau w_i^{(n)}- \beta(\tau)w_i^{(n)}
$$
With this notation it is possible to rewrite equation (\ref{eqmod})
as:
\begin{eqnarray}
\nonumber 
(1+\sum_{n=1}^{\infty}(w_j^{(n)}&+&w_k^{(n)}+\alpha_jk^{(n)}))
\sum_{n=1}^{n=\infty} P_i^{(n)}\\
& & = -\beta(\tau) \sum_{n=1}^{\infty}(
\frac{\alpha_{ij}^{(n)}+\alpha_{ik}^{(n)}}{2}+\frac{2}{3}\beta^{(n)})
\end{eqnarray} 

Keeping only terms of $n^{th}$ order in this equation gives:
$$
P_i^{(n)}=-\beta(\tau)(\frac{\alpha_{ij}^{(n)} +
  \alpha_{ik}^{(n)}}{2}+\frac{2}{3}\beta^{(n)}) 
+ \sum_{l=1}^{n-2} (w_j^{(l)}+w_k^{(l)}+\alpha_{jk}^{(l)})P_i^{(n-l)}
$$ 

$P_i^{(n)}$  can be computed using  only the $w_j^{(k)}$  with $k \leq
n-1$.   The differential  equations $d^2_\tau w_i^{(n)}  - \beta(\tau)
w_i^{(n)} = P_i^{(n)}$ can be recursively integrated to any order. The
fact that $P_i^{(n)}$ is of  order $(n)$ (except $P_i^{(1)}$ which  is
zero) is also  shown recursively.  If  we keep only  the linear growing
mode at first order, these equations can be  integrated very easily to
a high order.  Since equation  (\ref{eqmod}) is  exact only at  second
order the development will not be exact beyond that order.

However, as we have seen, the exact solution of equation (\ref{eqmod})
is much more accurate than the second order  solution and by developing
to a higher order we can have an analytical solution very close to the
numerically-integrated  one.   Another   interesting  feature of  this
development is that since  we can go  to  any order very easily  it is
possible to check if a  given quantity can be accurately  approximated
by perturbative theory. It  is possible that  some quantities  are not
well described by perturbative theory and are only computable with the
exact  solution.   For  example  $d^2_\tau w_i$   becomes  infinite at
collapse for the exact solution but  any perturbative calculation will
give a finite value.

\end{document}